\documentclass[camera]{jpaper}

\usepackage[nocompress]{cite}
\usepackage{algorithmic}
\usepackage{array}

\makeatletter
\let\MYcaption\@makecaption
\makeatother

\usepackage[font=footnotesize]{subcaption}

\makeatletter
\let\@makecaption\MYcaption
\makeatother

\usepackage{fixltx2e}
\usepackage{dblfloatfix}
\usepackage[nolessnomore, italic]{mathastext}
\usepackage[T1]{fontenc}
\usepackage[usenames,dvipsnames,svgnames,table]{xcolor}
\usepackage[normalem]{ulem}
\usepackage{enumitem}
\usepackage{setspace}
\usepackage{indentfirst}
\usepackage{footmisc}
\usepackage{fancyhdr}
\usepackage{authblk}
\usepackage[us,12hr]{datetime}
\usepackage[keeplastbox]{flushend}
\usepackage[hidelinks]{hyperref}

\usepackage{xspace}

\newboolean{publicversion}
\setboolean{publicversion}{true}

\ifthenelse{\boolean{publicversion}}{
  \pagenumbering{arabic}
  \newcommand{\grumbler}[2]{}
  \newcommand{\assign}[1]{}
  \newcommand{\respond}[3]{}
  \newcommand{\changesI}[0]{}
  \newcommand{\changesII}[0]{}
  \newcommand{\changesIII}[0]{}
  \newcommand{\changesIIII}[0]{}
  \newcommand{\changesIIIII}[0]{}
  \newcommand{\changesIIIIII}[0]{}
  
  \newcommand{\changesVIII}[0]{}

}{
  \pagenumbering{arabic}
  \newcommand{\grumbler}[2]{\textcolor{blue}{\bf #1: #2}}
  \newcommand{\assign}[1]{\textcolor{purple}{\bf RESPONSIBLE: #1}}
  \newcommand{\respond}[3]{\textcolor{#1}{\bf #2-response: #3}}
  \newcommand{\changesI}[0]{}
  \newcommand{\changesII}[1]{\textcolor{BrickRed}{#1}}
  \newcommand{\changesIII}[0]{}
  \newcommand{\changesIIII}[0]{}
  \newcommand{\changesIIIII}[0]{}
  \newcommand{\changesIIIIII}[0]{}

  \newcommand{\changesVIII}[1]{\textcolor{BrickRed}{#1}}
}

\newif\ifcameraready
\camerareadytrue
\newcommand{\versionnum}[0]{5}

\ifcameraready
\else
\fi

\ifcameraready
  \newcommand{\todo}[1][]{}
\else
  \newcommand{\todo}[1][]{\textbf{\fcolorbox{black}{red}{\color{white}{TODO}}} \underline{$\overline{\hbox{\emph{#1}}}$}}
\fi

\newcommand{\oursystem}[0]{\textit{Mosaic}\xspace}
\newcommand{\allocatorName}[0]{CoCoA\xspace}
\newcommand{\allocatorNameLong}[0]{\textit{\textbf{Co}ntiguity-\textbf{Co}nserving \textbf{A}llocation}\xspace}
\newcommand{\policyName}[0]{\textit{In-Place Coalescer}\xspace}
\newcommand{\compactionName}[0]{CAC\xspace}
\newcommand{\compactionNameLong}[0]{\textit{\textbf{C}ontiguity-\textbf{A}ware \textbf{C}ompaction}\xspace}
\newcommand{\titleShort}[0]{\oursystem}

\newcommand{\paragraphbe}[1]{{\textbf{#1}} }

\newcommand*\mycirc[1]{%
  \tikz[baseline=(char.base)]{
  \node[shape=circle,draw,inner sep=0.5pt,fill=black,text=white,font=\bfseries] (char) {#1};}%
}%

\newcommand*\mycirctwo[1]{%
  \tikz[baseline=(char.base)]{
  \node[shape=circle,draw,inner sep=0.0pt,fill=black,text=white,font=\bfseries] (char) {#1};}%
}%

\newcommand{\ignore}[1]{}

\begin{document}
\title{Mosaic: An Application-Transparent\\Hardware--Software Cooperative Memory Manager for GPUs}

\newcommand{\authspace}[0]{\hspace{18pt}}
\newcommand{\affilspace}[0]{\hspace{15pt}}

\author{%
    Rachata Ausavarungnirun$^1$\qquad%
    Joshua Landgraf$^2$\qquad%
    Vance Miller$^2$\qquad%
    Saugata Ghose$^1$%
    \vspace{2pt}\\ %\authspace
    Jayneel Gandhi$^3$\qquad%
    Christopher J. Rossbach$^{2,3}$\qquad%
    Onur Mutlu$^{4,1}$}
\affil{%
{\it%
	$^1$Carnegie Mellon University\qquad%
	$^2$The University of Texas at Austin%
}\vspace{2pt}\\{\it%
	$^3$VMware Research\qquad%
    $^4$ETH Z{\"u}rich}}

%\author{%
%{Saugata Ghose\affilCMU}%
%\qquad%
%{Yixin Luo\affilCMU}%
%\qquad%
%{Onur Mutlu\affilETH\affilCMU}%
%\vspace{3pt}\\%
%{\it\affilCMU Carnegie Mellon University \qquad \affilETH ETH Z{\"u}rich}%
%\vspace{3pt}%
%}

\maketitle

% !TEX root=../paper.tex

\begin{abstract}

This paper summarizes the idea and key contributions of Mosaic, which was
published at MICRO 2017~\cite{mosaic},
and examines the work's significance and future potential.
Contemporary discrete GPUs support rich memory management features
such as virtual memory and demand paging. 
These features simplify GPU programming by providing
a virtual address space abstraction similar to CPUs and eliminating
manual memory management, 
but they introduce high performance \changesI{overheads} during
(1)~address translation and (2)~page faults.
A GPU relies on high degrees of \emph{thread-level parallelism} (TLP) to hide 
memory latency.
Address translation can undermine TLP, as a single miss in the translation
lookaside buffer (TLB) invokes an expensive serialized page table walk that 
often stalls \emph{multiple} threads.  
% Due to the limited TLB capacity,
% these TLB misses occur more frequently as application data sets grow, and as
% multiple applications interfere with each other.
\changesI{Demand paging} can also undermine TLP, as \changesI{multiple threads often stall while they wait for an}
expensive data transfer over the \changesI{system I/O (e.g., PCIe) bus} when the GPU demands a page.
%over the PCIe bus when the GPU demands a page.

% Unfortunately, the features introduce high
% overheads for (1)~address translation and (2)~page faults (which require 
% expensive data transfers over PCIe), which can cripple GPU performance.}
% % These features dramatically
% % simplify GPU programming by providing a familiar virtual address space
% % abstraction, and eliminating manual memory management, but they come at the
% % cost of additional overheads for address translation and
% % page-fault-driven PCIe transfers. The overheads can be crippling.

In \changesI{modern} GPUs, \changesI{we face} a trade-off \changesI{on how
the page size used for memory management affects address translation and demand paging.}
% \changesI{when we try to reduce} these two overheads.
\changesI{The address translation overhead is lower when we employ a \emph{larger} page size}
(e.g., 2MB \emph{large pages}, compared with
conventional 4KB \changesI{\emph{base pages}}), which \changesI{increases} TLB coverage and thus \changesIII{reduces}
TLB misses.
Conversely, \changesI{the} demand paging \changesI{overhead is lower when we
employ a \emph{smaller} page size}, which \changesI{\changesI{decreases} the \changesIII{system I/O bus} transfer latency}.
%and improve page prefetching.
% is dominated by transfer latencies,
% which are best mitigated by overlapping transfer with kernel execution,
% and by predictive prefetching; both strategies are better served
% by smaller page sizes.
% %GPU kernel threads, and avoided where possible using techniques
% %such as predictive prefetch. Larger pages magnify load-to-use
% %intervals with longer PCIe transfers, increasing the average number
% %of threads stalled by a given fault. Demand-paging for discrete
% %GPUs is best served by smaller pages; indeed previous work has
% %dismissed huge pages as a viable alternative for these reasons.
%We find that GPGPU applications present an opportunity to exploit this trade-off, 
\changesI{Support for \emph{multiple page sizes} can help relax the page size
trade-off so that address translation and demand paging optimizations work
together synergistically.  However, existing \changesIII{page} \emph{coalescing}
(i.e., merging base pages into \changesIIIII{a large page}) and \emph{splintering} (i.e., splitting a
large page into base pages) policies require costly base page migrations that
\changesIII{undermine the benefits multiple} page sizes provide.}
\changesI{In this paper, we observe} that GPGPU applications present an opportunity 
% \changesI{for address
% translation and demand paging optimizations to} work together \changesI{synergistically}, 
\changesI{to support multiple page sizes without costly data migration,}
as \changesI{the applications} perform most of their memory allocation \changesI{\emph{en masse}
(i.e., \changesI{they allocate a large number of base pages at once})}.
% \changesI{We show that} with \changesI{intelligent} memory allocation strategies, 
\changesI{We show that this \emph{en masse} allocation allows us to
create intelligent memory allocation policies which ensure that base pages that are contiguous
in virtual memory are allocated to contiguous physical memory pages.
As a result, coalescing and splintering operations no longer
need to migrate base pages.}
% \changesI{a memory manager
% can effectively} coalesce multiple base pages into a single large page without the need
% to \changesI{perform costly data migration} most of the time.

We introduce \titleShort, a GPU \changesI{memory manager} that
provides \emph{application-transparent} \changesI{support for multiple page sizes}.
\titleShort uses base pages to transfer data over \changesI{the system I/O bus}, and allocates physical
% \titleShort transfers data over I/O bus using small page granularity, and allocates physical
memory in a way that \changesI{(1)~}preserves \changesI{base page contiguity} 
and \changesI{(2)~}ensures that a 
large page \changesIIII{frame} contains \changesI{pages from only} a single memory protection domain.
\changesI{We take advantage of this allocation strategy to design
\changesI{a novel in-place page \changesIII{size selection mechanism} that avoids data migration.}
%\changesI{a novel in-place page \changesIII{size selection mechanism} that avoids data migration.}
%  novel \emph{coalescing} 
% (i.e., merging base pages into a large page) and \emph{splintering} (i.e. splitting a
% large page into base pages) primitives that can be implemented as low-latency
% atomic operations. 
\changesI{This \changesIII{mechanism} allows} the TLB to use large \changesI{pages,} 
reducing address translation overhead.
% while splintering operations allow 
\changesIII{During data transfer, this mechanism enables the GPU}
to transfer \changesI{\emph{only}} the base pages that are needed by the application over the
system I/O bus, \changesI{keeping demand paging overhead low}.}
% significantly increase the TLB
% reach through synchronization-free page coalescing and splintering primitives.
% \changesI{the apparent tension between address translation overheads 
% and demand paging performance by providing \emph{transparent}} huge page
% support for discrete GPUs. \changesI{\titleShort transparently and dynamically
% changes virtual memory mappings during workload execution, by using small 4KB
% pages to perform demand paging when transferring data over PCIe, while
% employing fast address translation based on 2MB large pages to significantly
% reduce the TLB miss rate.}
% % supports fast
% % address translation based on huge pages, with the demand
% % performance enabled by small pages, transparently and dynamically
% % changing virtual memory mappings in response to workloads.
% \titleShort tracks spatial and temporal reference locality
% to inform \changesI{a \emph{contiguity-preserving}} memory management policy,
% which reduces the need for compaction-induced memory 
% to create contiguity.
% \titleShort uses two novel mechanisms, \allocatorName and \policyName, to 
% implement synchronization-free coalesce and splinter primitives
% with the performance cost of a single atomic pointer swap.
Our evaluations show that \titleShort reduces address translation
overheads \changesI{while efficiently achieving the benefits of demand paging},
% near-negligible address translation overheads
% with minimal impact on demand paging performance, 
compared to a contemporary GPU that uses only a 4KB page size.
Relative to \changesI{a} state-of-the-art GPU \changesI{memory manager}, 
\titleShort improves the performance \changesI{of homogeneous and 
heterogeneous multi-application workloads by 55.5\% and 29.7\% on average, 
respectively, coming within 6.8\% and 15.4\% of the performance 
\changesI{of an ideal TLB where all TLB requests are hits}.}
\end{abstract}

% !TEX root=../paper.tex

\section{Introduction}
\label{sec:intro}

\sloppypar{Graphics Processing Units (GPUs) are used for an ever-growing
range of application domains due to their capability to provide high throughput.
\changesI{GPUs provide a high amount of throughput but \changesII{they require a different} programming
model \changesII{than CPUs}, making \changesII{their general} adoption difficult.}
Recent support within GPUs for \emph{memory virtualization} features,
such as a unified virtual address space~\cite{fermi,huma}, demand
paging~\cite{pascal}, and preemption~\cite{pascal,firepro}, can ease
programming by allowing developers to exploit key benefits 
such as application portability and multi-application execution. }
Hardware-supported memory virtualization relies on address translation to map
each virtual memory address to a physical address within the GPU memory. 
Address translation uses page-granularity virtual-to-physical mappings that are 
stored within a multi-level \emph{page table}.  To look up
a mapping within the page table, the GPU performs a page table \emph{walk},
where a page table walker traverses through each level of the page table in
main memory until the walker locates the \emph{page table entry} for the
requested mapping in the last level of the table.
GPUs with virtual memory support \changesIII{have} \emph{translation lookaside buffers} 
(TLBs), which cache page table entries and avoid the need to perform a page
table walk for the cached entries, \changesII{thereby} reducing the address translation latency.

State-of-the-art GPU memory virtualization provides support for \emph{demand 
paging}~\cite{powers-hpca14,tianhao-hpca16,cc-numa-gpu-hpca15,huma,pascal}. %~\cjr{check pls}.
In demand paging, all of the memory used by a GPU application does \emph{not} need to
be transferred to \changesI{the} GPU memory at the beginning of application execution.  Instead,
during application execution, when a GPU thread issues a memory request to a
page that has not yet been allocated in \changesI{the} GPU memory, the GPU issues a \emph{page
fault}, at which point the data for that page is transferred over the off-chip system I/O bus 
(e.g., the PCIe bus~\cite{pcie} in contemporary systems)
from the CPU memory to the GPU memory. The transfer
requires a long latency due to its use of an off-chip bus.  Once the transfer
completes, the GPU runtime allocates a physical GPU memory \changesI{address} to the page,
and the thread can complete its memory request.

\paragraphbe{GPU Virtualization Challenges.}
Two fundamental challenges prevent further adoption of virtualization
in GPUs: (1)~the address translation
challenge, and (2)~the demand paging challenge.
The address translation challenge stems from a long latency process that consists of
a series of \emph{serialized} memory accesses required to traverse the page
table~\cite{powers-hpca14, pichai-asplos14}.  As many threads can \changesII{access different data present} in
a single page, these serialized page walk accesses significantly limit GPU
\changesII{concurrency}, by lowering \emph{thread-level parallelism} (TLP) and
\changesII{thereby reducing the} latency hiding capability of a GPU.  \emph{Translation
lookaside buffers} (TLBs) can reduce the latency of address translation by
caching recently-used address translation information.  Unfortunately, as
application working sets and DRAM capacity have increased in recent years,
state-of-the-art TLB designs~\cite{powers-hpca14,pichai-asplos14} suffer \changesII{from}
\emph{poor TLB reach}, i.e., the TLB covers only a small fraction of the
physical memory working set of an application.  We found that the poor TLB
reach has a detrimental effect on \changesII{GPU performance}, because a \emph{single} TLB miss can
stall \emph{hundreds} of threads at once, undermining TLP within a GPU and
significantly reducing
performance~\cite{abhishek-ispass16,gpu-arch-microbenchmarking,mosaic}. 

\begin{figure}[t]
\centering
\vspace{5pt}
%  \vspace{-0.5em}% 
  \includegraphics[width=\columnwidth]{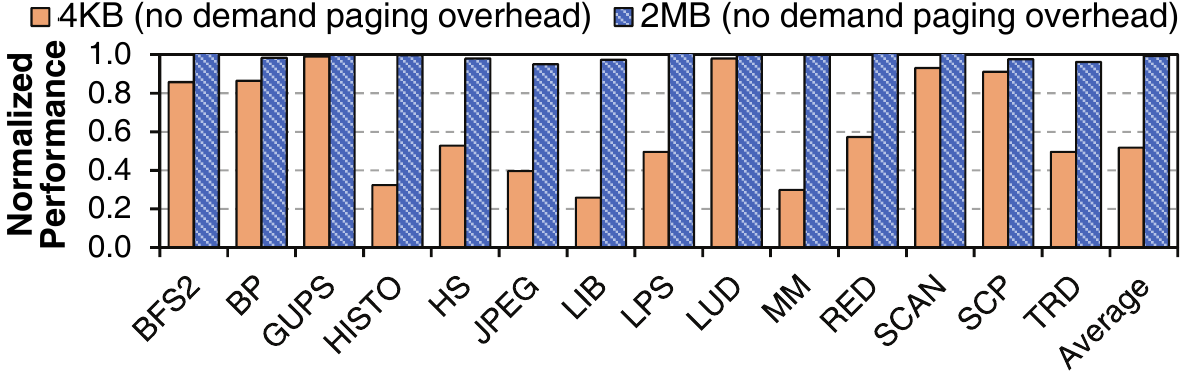}%
%  \vspace{-1em}%
  \caption{\changesI{Performance of a GPU with \emph{no demand paging
   overhead}, using (1)~4KB base pages and
   (2)~2MB large pages,
   normalized to the performance of a GPU with an ideal TLB}. Reproduced from~\cite{mosaic}.}%
  \label{fig:mask-summary}%
%  \vspace{-0.5em}%
\end{figure}

Figure~\ref{fig:mask-summary} shows the performance of two GPU-MMU designs:
(1)~a design that uses the base 4KB page size, and (2)~a design that uses a 2MB
large page size, \changesI{where both designs have} \emph{no demand paging
overhead} (i.e., \changesI{the} system I/O bus transfer takes zero cycles
\changesIII{to transfer a page}). We normalize the performance of the two
designs to \changesI{a GPU with an ideal TLB,} where \emph{all} TLB requests
hit in the L1 TLB. \changesII{Our full experimental methodology is described in detail in our MICRO 2017 paper~\cite{mosaic}}. 
\changesI{We make two \changesII{major} observations from the figure.}

%%%% TODO: Y-axis label

\changesI{First, compared to the ideal TLB, %where all TLB requests hit in the L1 TLB, 
the GPU-MMU with 4KB base pages experiences
an average performance \changesIII{loss} of 48.1\%. % \sg{check this!}
We observe that with 4KB base pages,
% optimization of the memory subsystem can minimize the
% overhead lost to address translation through interference. However, cases 
a single TLB miss often stalls \emph{many} of the warps, which undermines
the latency hiding behavior of the SIMT execution model used by GPUs.}
\changesI{Second, the figure shows that} 
% for the design with 2MB pages without any demand paging overhead,
% % (i.e., we assume that no far-faults occur),
% we observe that 
using a 2MB page size with the same number of TLB entries 
as the 4KB design \changesI{allows applications to come within 2\% of the 
ideal TLB performance.  We find that with 2MB pages, the TLB has a much larger
reach, which
reduces the TLB miss rate substantially.}
% With 2MB pages, the TLB has a much larger reach, which allows applications to come
% within 2\% of the \changesI{all-hit TLB} performance (i.e., where \emph{all} TLB 
% accesses hit in the L1 TLB). 
Thus, there is strong incentive to use large
pages for address translation.

To increase the TLB reach,
\emph{large pages} (e.g., the 2MB or 1GB pages \changesII{used in many} modern CPU \changesII{architectures}~\cite{haswell,skylake})
can be employed. However, large pages increase the risk of \emph{internal fragmentation}, where
a portion of the large page is unallocated (or unused). Internal fragmentation occurs because it \changesII{might often be}
difficult for an application to \emph{completely} utilize large contiguous regions of memory.
This fragmentation leads to (1)~\emph{memory bloat},
where a much greater amount of physical memory is allocated 
than the amount of memory that the application needs;
and (2)~longer memory access latencies,
due to a \changesII{potentially} lower effective TLB reach and more page faults~\cite{ingens}.

The demand paging challenge stems from a \emph{page fault}, which requires a
long-latency data transfer for an entire page over the system I/O
bus~\cite{pcie}.
\changesII{Since} GPU threads often access data in the same page due to data locality,
a single page fault can cause \emph{multiple} threads to stall at once.
As a result, the page fault can significantly reduce the amount of TLP that 
the GPU can exploit, and \changesII{thus} significantly degrade performance~\cite{tianhao-hpca16,mosaic}.

Unlike address translation, which benefits from \emph{larger} pages, the
overhead of demand paging is smaller when \emph{a smaller page size} is used.
A larger \changesII{amount of} data transfer increases the transfer time, increases the amount of time
that GPU threads stall, and decreases TLP. Furthermore, as the size of a page
increases, there is a greater probability that an application does \emph{not} need all
of the data in the page. As a result, threads may stall for a longer time
without gaining any further benefit in return. Based on these two conflicting
observations, memory virtualization in GPU systems has a fundamental trade-off due
to the page size choice. We provide more detail on the trade-off in our \changesII{MICRO 2017} paper~\cite{mosaic}.

\section{Mosaic}

\begin{figure}[h!]%
\centering
	\includegraphics[width=\columnwidth]{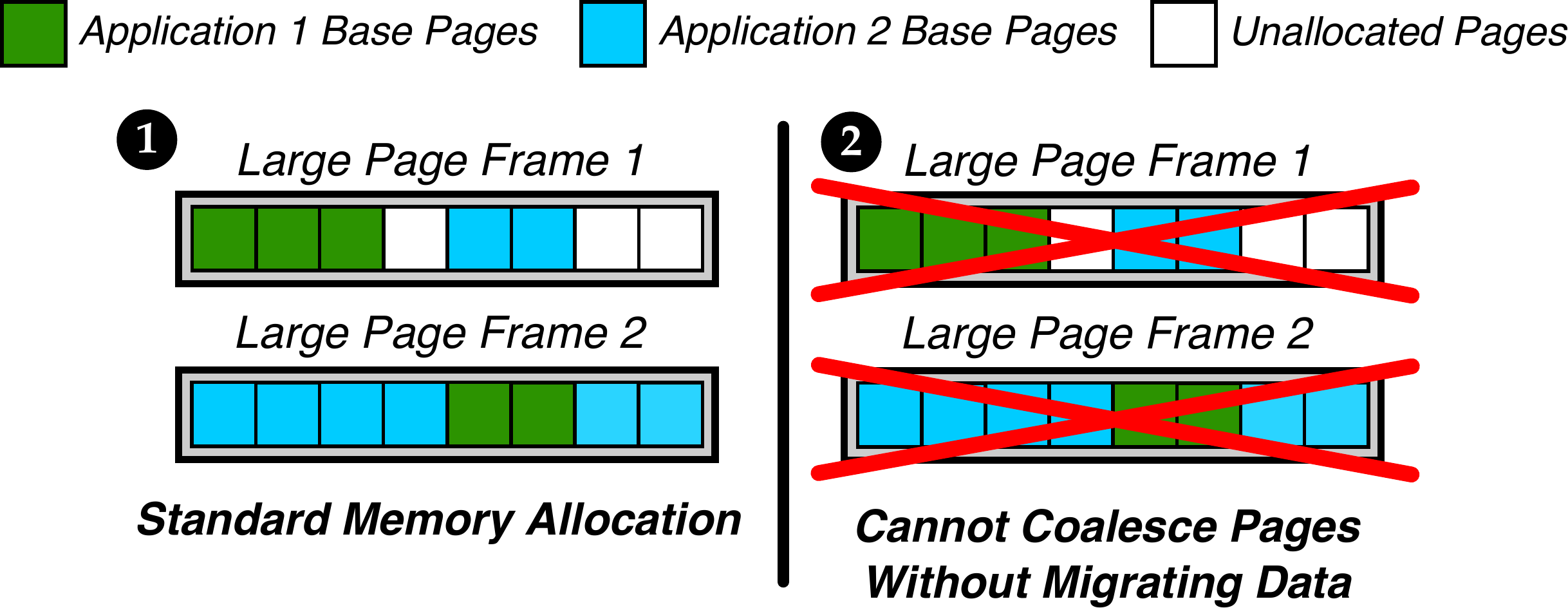}%
\caption{Page allocation and coalescing behavior of a state-of-the-art GPU memory manager~\cite{powers-hpca14}. Adapted from~\cite{mosaic}.}
	\label{fig:tlb-mosaic-intro-base}
\end{figure}

\begin{figure}[h!]%
\centering
	\includegraphics[width=0.793\columnwidth]{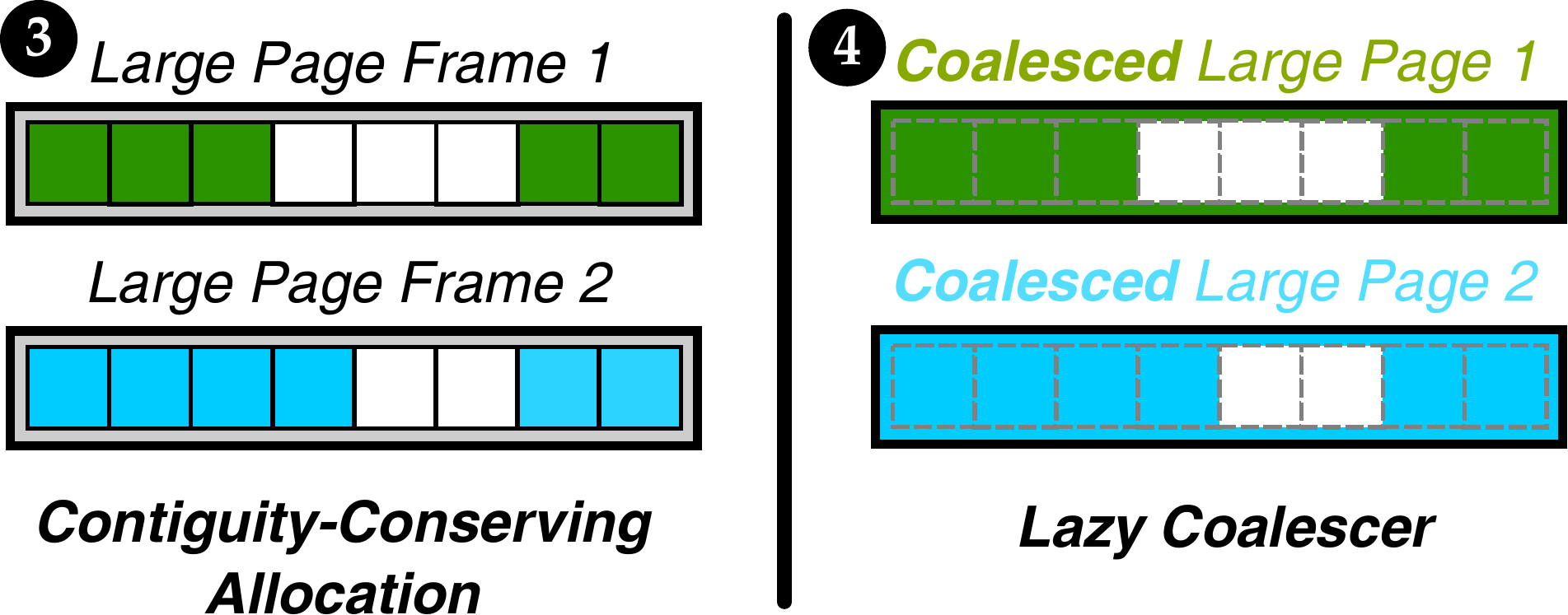}%
\caption{Page allocation and coalescing behavior of \titleShort. Adapted from~\cite{mosaic}.}
	\label{fig:tlb-mosaic-intro}
\end{figure}

In our MICRO 2017 paper~\cite{mosaic}, \changesI{we propose \titleShort, \changesII{a new GPU memory management scheme that aims to get
the best of both small and large page sizes}.} \titleShort relaxes the page size
trade-off by using \emph{multiple} page sizes \emph{transparently} to the
application, and, thus, to the programmer.  With multiple page sizes, and the
ability to change virtual-to-physical mappings dynamically, the GPU system can
support \emph{good TLB reach} by using large pages for address translation, while
providing \emph{good demand paging performance} by using base pages for data
transfer. However, while coalescing multiple small pages into a large page requires a
contiguous region, \changesII{existing memory} allocation mechanisms make it difficult to find regions
of physical memory where base pages can be coalesced without a large number of page
migration operations. \changesII{This is because existing GPU
memory allocation mechanisms do not allocate base pages in a manner that is aware of the
contiguity of memory allocated to each application.}
Figure~\ref{fig:tlb-mosaic-intro-base} shows how a state-of-the-art GPU memory
manager~\cite{powers-hpca14} allocates memory for two applications.  Within a
single \emph{large page frame} (i.e., a contiguous piece of physical memory
that is the size of a large page and whose starting address is page aligned),
the GPU memory manager allocates base pages from both Applications~1 and 2
(\mycirc{1} in Figure~\ref{fig:tlb-mosaic-intro-base}).  As a result, the memory manager \emph{cannot}
coalesce the base pages into a large page (\mycirc{2}) without first migrating
some of the base pages, which would incur a high latency.
Instead, \titleShort allocates physical base pages in a way that avoids the need to migrate data 
during coalescing (\mycirc{3} in Figure~\ref{fig:tlb-mosaic-intro}), and uses a simple coalescing mechanism to 
combine base pages into large pages (e.g., 2MB) and thus increase TLB reach (\mycirc{4} in Figure~\ref{fig:tlb-mosaic-intro}).


%\begin{figure}[h!]%
%\centering
%\vspace{-15pt}%
%\subfloat[State-of-the-art GPU memory management~\cite{powers-hpca14}.]{{
%	\includegraphics[width=\columnwidth]{figs/mosaic-intro-base.pdf}%
%	\label{fig:tlb-mosaic-intro-base}
%}}%
%\vspace{-5pt}
%\subfloat[Memory management with \titleShort.]{{
%	\includegraphics[width=0.793\columnwidth]{figs/mosaic-intro-sample.pdf}%
%	\label{fig:tlb-mosaic-intro}
%}}%
%\vspace{-8pt}%
%\caption{Page allocation and coalescing behavior \changesI{of GPU memory managers: 
%(a)~state-of-the-art~\cite{powers-hpca14}, (b)~\titleShort}. Reproduced from~\cite{mosaic}.}%
%\vspace{-0.7em}
%\end{figure}

We make a \emph{key observation} about the memory behavior of
contemporary general-purpose GPU (GPGPU) applications. \changesII{We find that} the vast majority of
memory allocations in GPGPU applications are performed \emph{en masse} (i.e., a large
number of pages are allocated at the same time).  The \emph{en masse} memory
allocation presents us with an opportunity: with so many pages being allocated
at once, we can rearrange how we allocate the base pages to ensure that
(1)~\emph{all} of the base pages allocated within a large page frame belong to the
\emph{same} virtual address space, and
(2)~base pages that are contiguous in virtual memory are allocated to a
contiguous portion of physical memory and aligned within the large page frame.

\begin{figure*}[ht!!!]
  \centering
%  \vspace{-0.5em}%
  \includegraphics[width=1.9\columnwidth]{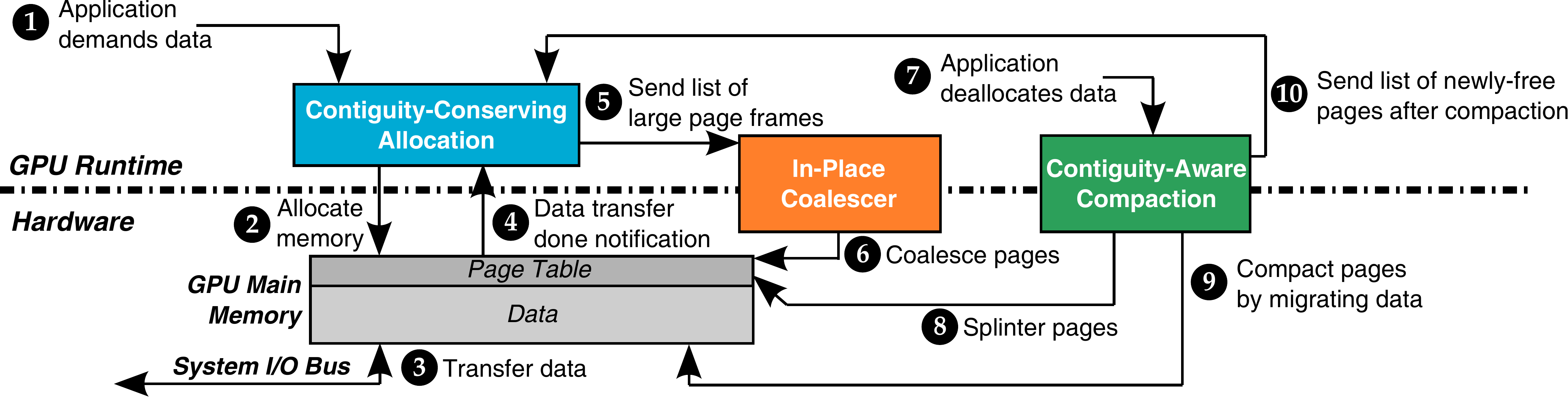}%
%  \vspace{-1em}%
  \caption{High-level overview of \titleShort, showing how and when its three 
  components interact with the GPU memory. Reproduced from~\cite{mosaic}.}
%\vspace{-1em}
  \label{fig:mosaic-overall}
\end{figure*}

%To enable an application-transparent multiple page size support in GPUs,
\sloppypar{\titleShort is designed \changesII{to achieve these two goals. It consists of} 
three major components: \allocatorNameLong
(\allocatorName), the \policyName, and \compactionNameLong (\compactionName).
These three components work together to \emph{coalesce} (i.e., combine) base 
pages \changesII{into large pages and \emph{splinter} (i.e., split apart)
large pages back to base pages} during memory management.
Memory management operations for \titleShort take place at two times: (1)~when
memory is \emph{allocated}, and (2)~when memory is \emph{deallocated}.
\changesII{We describe what happens at each component briefly. Figure~\ref{fig:mosaic-overall}
depicts the three components of \titleShort, and we will use Figure~\ref{fig:mosaic-overall}
to provide a walkthrough of the actions taken during memory allocation and deallocation.}
}

% Rachata -- Need to edit the part below. 

\paragraphbe{Memory Allocation.}
When a GPGPU application wants to access data that is \emph{not} currently in the GPU
memory, it sends a request to the GPU runtime (e.g., OpenCL, CUDA runtimes) to
transfer the data from the CPU memory to the GPU memory (\mycirc{1} in 
Figure~\ref{fig:mosaic-overall}).
A GPGPU application typically allocates a large number of base pages at the same time.
\allocatorName allocates space within the GPU memory (\mycirc{2})
for the base pages, \changesIII{working to conserve the contiguity of base 
pages\changesIIII{,} if possible during allocation}.  Regardless of contiguity, 
\allocatorName provides a \emph{soft guarantee} that a single large 
page frame contains base pages from only a \emph{single} application.
Once the base page is allocated, \allocatorName initiates the data transfer 
across the system I/O bus (\mycirc{3}).
When the data transfer is complete (\mycirc{4}), \allocatorName notifies 
the \policyName that allocation is done \changesIIII{by sending a list of the large page
frame addresses that were allocated} (\mycirc{5}).
For each \changesIIII{of these large page frames}, the \changesIII{runtime portion of the}
\policyName then checks to see whether 
(1)~\emph{all} base pages \changesIIIII{within the large page frame} have been allocated, and
(2)~\changesIII{the base pages within the large page frame are contiguous in
both virtual and physical memory}.
If both conditions are true, the \changesIII{hardware portion of the \policyName 
updates} the page table to coalesce the base pages into a large page 
\changesIII{(\mycirc{6})}. \changesII{Section 4.3 of our MICRO 2017 paper~\cite{mosaic}
describes how page tables are modified to support coalescing.}

\paragraphbe{Memory Deallocation.}
When a GPGPU application \changesII{would like} to deallocate memory (e.g., when an application
kernel finishes), it sends a deallocation request to the GPU runtime (\mycirc{7}).
For all deallocated base pages that are coalesced into a large page, the runtime 
invokes \compactionNameLong (\compactionName) for the corresponding large page.  \changesIII{The runtime portion of}
\compactionName checks to see whether the large page has a high degree of
\emph{internal fragmentation} (i.e., if the number of unallocated base pages
within the large page exceeds a predetermined threshold).
For each large page with high internal fragmentation, \changesIII{the hardware
portion of \compactionName updates} the
page table to splinter the large page back into its constituent base pages \changesIII{(\mycirc{8})}.
Next, \compactionName compacts the splintered large page frames, by migrating 
data from multiple splintered large page frames into a single large page frame
(\mycirc{9}).  Finally, \compactionName notifies \allocatorName of the large
page frames that are now free after compaction (\mycirctwo{10}), which 
\allocatorName can use for future memory allocations. \changesII{We describe each
component of \titleShort in more detail in Sections~4.2, 4.3, and 4.4 of our MICRO 2017
paper~\cite{mosaic}.} % and provide the implementation of the corresponding components in Sections 5~\cite{mosaic}.}

%The design of \titleShort provides an application-transparent GPU memory
%manager that \emph{efficiently} supports multiple page sizes. \titleShort uses a
%simple but novel mechanism to allocate contiguous virtual pages to contiguous
%physical pages in the GPU memory, and exploits this property to coalesce
%contiguously-allocated base pages into a large page for address translation
%with low overhead and no data migration. \titleShort minimizes the overhead of
%demand paging by only using base pages during paging.  We show that \titleShort
%effectively improves TLB reach while efficiently achieving the benefits of
%demand paging.  Overall, \titleShort improves the average performance of
%homogeneous and heterogeneous multi-application workloads by 55.5\% and 29.7\%,
%respectively, over a state-of-the-art GPU memory manager~\cite{powers-hpca14}.

\section{Evaluation Methodology}

Table~\ref{table:config} \changesIII{shows the system \changesIIII{configuration} we simulate
for our evaluations, \changesIIIII{including the configurations of the GPU cores and memory partitions.}
%(see Section~\ref{sec:bkgd:gpu})}.}
\changesI{We modify the MAFIA framework~\cite{mafia}, which uses GPGPU-Sim 3.2.2~\cite{gpgpu-sim},
to evaluate \titleShort}. We add a memory allocator into
cuda-sim, the CUDA simulator within GPGPU-Sim, to handle all 
virtual-to-physical address translations and to provide memory protection.
We add an accurate model of address translation to GPGPU-Sim, including TLBs,
page tables, and a page table walker.  The page table walker is shared across
all SMs, and allows up to 64~concurrent walks.  Both the L1 and
L2 TLBs have separate entries for base pages and large 
pages~\cite{rmm, jayneel-isca16, karakostas.hpca16, binh-colt, binh-micro15, prediction-tlb}.
Each TLB contains miss status holding registers \changesIIIIII{(MSHRs)~\cite{kroft-isca81}} to track in-flight 
page table walks.
Our simulation infrastructure supports demand paging by detecting page faults
and faithfully modeling the system I/O bus (i.e., PCIe) latency
based on measurements from NVIDIA GTX 1080 cards~\cite{nvidia-1080}.
% We evaluate GPGPU applications from the Parboil~\cite{parboil}, SHOC~\cite{shoc},
% LULESH~\cite{lulesh,lulesh-origin}, Rodinia~\cite{rodinia}, and CUDA
% SDK~\cite{cuda-sdk} suites.
%\footnote{Our 
%experience with the NVIDIA GTX 1080 suggests that production \changesIIIII{GPUs} 
%perform significant prefetching to reduce latencies when reference patterns 
%are predictable. This feature is not modeled \changesIII{in our simulations}.}
We use a worst-case model for the performance of our compaction mechanism
conservatively, by stalling the \emph{entire} GPU \changesIII{(all SMs)} and 
flushing the pipeline.  
\changesII{We have publicly released our simulator modifications as open source
software~\cite{mosaic.github, safari.github}.}

%More details about our modifications can be found in
%our extended technical report~\cite{mosaic-tech}.

\begin{table}[ht!]
\begin{footnotesize}
\centering
%\vspace{-0.5em}
\begin{tabular}{ll}
         \cmidrule(rl){1-2} 
%        \hline
%\textbf{System Overview}           &  30 cores, 6 memory partitions\\
%        \cmidrule(rl){1-2}
\multicolumn{2}{c}{\textbf{\changesIII{GPU Core Configuration}}} \\ 
        \cmidrule(rl){1-2}\morecmidrules\cmidrule(rl){1-2}
\textbf{Shader Core}           &  30 cores\changesIII{,} 1020 MHz, GTO warp scheduler~\cite{ccws}\\
\textbf{Config}           &  \\
        \cmidrule(rl){1-2}
\textbf{Private L1 Cache}    &  16KB, 4-way associative, LRU, L1 misses are \\ & coalesced before accessing L2, 1-cycle latency \\
        \cmidrule(rl){1-2} 
\textbf{Private L1 TLB}    &  \changesIIIII{128 base page/16 large page} entries per core,\\
 & fully associative, LRU, single port, 1-cycle latency \\ 
        \cmidrule(rl){1-2}
%& \\
\cmidrule(rl){1-2}
\multicolumn{2}{c}{\textbf{\changesIII{Memory Partition Configuration}}}\\
\multicolumn{2}{c}{\changesIIIII{(6 memory partitions in total}} \\
\multicolumn{2}{c}{\changesIIIII{with each partition accessible by all 30~cores)}} \\
        \cmidrule(rl){1-2}\morecmidrules\cmidrule(rl){1-2}
\textbf{Shared L2 Cache}   &  2MB total, 16-way associative, LRU, 2 cache banks, \\ & 2 ports per memory partition, 10-cycle latency \\
        \cmidrule(rl){1-2} 
\textbf{Shared L2 TLB}   &  \changesIIIII{512 base page/256 large page} entries,\\ & 16-way/fully-associative (base page/large page),\\ & , \changesIIII{non-inclusive, LRU,}\changesIII{2 ports,} 10-cycle latency \\
        \cmidrule(rl){1-2} 
        \textbf{DRAM}   & 3GB GDDR5~\cite{gddr5,kim-cal2015}, 1674 MHz,\\ & 6 channels, 8 banks per rank, \\ & FR-FCFS scheduler~\cite{fr-fcfs,frfcfs-patent}, burst length 8\\
         \cmidrule(rl){1-2} 
    \end{tabular}%
\caption{Configuration of the simulated system. Adapted from~\cite{mosaic}.}
%\vspace{-2.5em}
  \label{table:config}%
\end{footnotesize}%
\end{table}%

\changesII{
We evaluate the performance of \titleShort using both \emph{homogeneous} and 
\emph{heterogeneous} workloads.
We categorize each workload based on the number of 
concurrently-executing applications, which ranges from one to five for our
homogeneous workloads, and from two to five for our heterogeneous workloads.
We form our homogeneous workloads using multiple copies of the same
application. We build 27~homogeneous workloads for each category 
using GPGPU applications from the Parboil~\cite{parboil}, SHOC~\cite{shoc},
LULESH~\cite{lulesh,lulesh-origin}, Rodinia~\cite{rodinia}, and CUDA
SDK~\cite{cuda-sdk} suites. We form our heterogeneous workloads by randomly selecting a
number of applications out of these 27~GPGPU applications.
We build 25 heterogeneous workloads per category.
In total we evaluate 235~homogeneous and heterogeneous workloads.}

We compare \titleShort to two mechanisms:
(1)~\emph{GPU-MMU}, a \changesIIII{baseline GPU with a state-of-the-art memory manager} based \changesIIII{on}
the work by Power et al.~\cite{powers-hpca14}; %, which we explain in detail in 
%Section~\ref{sec:baseline-eval}; 
and (2)~\emph{Ideal TLB}, a GPU with an ideal TLB, where every address 
translation request hits in the L1 TLB (i.e., there are no TLB misses).
We report workload performance using the weighted speedup metric~\cite{harmonic_speedup,ws-metric2},
which is calculated as:
\vspace{5pt}%
\begin{equation}
    \text{Weighted Speedup} = \sum{\frac{IPC_{shared}}{IPC_{alone}}}
\vspace{5pt}%
\end{equation}
where $IPC_{alone}$ is the \changesII{retired instructions per cycle} (IPC) of an application in the workload that runs on
the same number of shader cores using the baseline state-of-the-art
configuration~\cite{powers-hpca14}, but does \emph{not} share GPU resources with any
other applications; and $IPC_{shared}$ is the IPC of the application when it
runs concurrently with other applications.  We report the performance of each
application within a workload using IPC.

\changesII{Section 5 of our MICRO 2017 paper~\cite{mosaic} provides more detail on our experimental
methodology}.

%%%%% HERE

\section{Experimental Results}

Figure~\ref{fig:homo-mosaic-eval} shows the performance of \titleShort for the
homogeneous workloads \changesII{we evaluated}.  
We make two observations from the figure.
First, we observe that \titleShort{} is able to recover most of the performance lost due to
the overhead of address translation (i.e., an ideal TLB) in homogeneous workloads.  
Compared to the GPU-MMU baseline, \titleShort improves \changesII{system} performance by 55.5\%,
averaged across all 135 of our homogeneous workloads.
The performance of \titleShort comes within 6.8\% of the \emph{Ideal TLB} performance, 
indicating that \titleShort is effective at extending the TLB reach.
%  without negatively harming demand paging performance.
Second, we observe that \titleShort provides good scalability.
As we increase the number of concurrently-executing applications, \changesII{which puts more pressure on the shared TLBs,}
we observe that the performance of \titleShort remains close to
the \emph{Ideal TLB} performance.

\begin{figure}[h]
\centering
% \vspace{0.5em}%
  \includegraphics[width=\columnwidth]{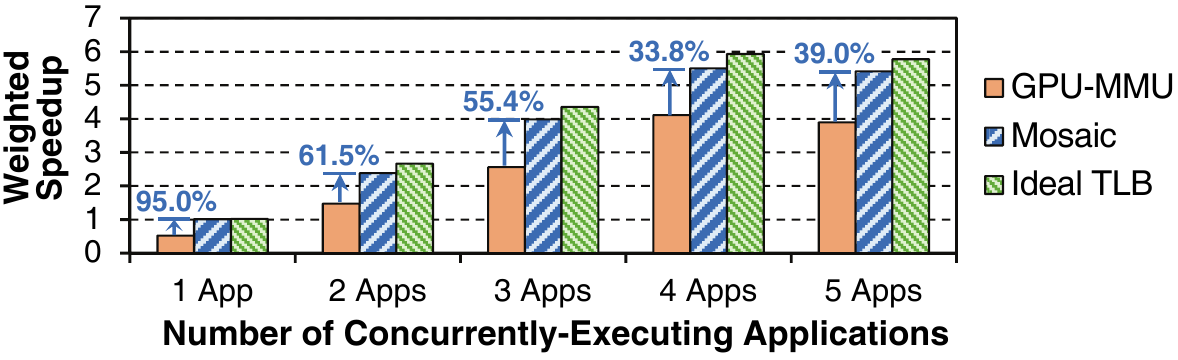}%
%  \vspace{-1em}%
  \caption{Homogeneous workload performance of GPU memory managers as
we vary the number of concurrently-executing applications in each workload. Reproduced from~\cite{mosaic}.}%
%  \vspace{-0.5em}%
  \label{fig:homo-mosaic-eval}
\end{figure}

Figure~\ref{fig:mosaic-eval} shows the performance of \titleShort for
heterogeneous workloads that consist of multiple \changesII{different} randomly-selected 
GPGPU applications \changesII{(100 workloads in total)}. From the figure, we observe that on average across all
of the workloads, \titleShort{}
provides a performance improvement of 29.7\% over GPU-MMU, and
comes within 15.4\% of the \emph{Ideal TLB} performance. We find that the
improvement comes from the significant reduction in the TLB miss 
rate with \titleShort. \changesII{We also see that \titleShort's scalability
is good, as the number of applications increases, yet there is still room
for improvement  to reach the performance of \emph{Ideal TLB}. We conclude
that \titleShort is a more effective memory manager than the state-of-the-art.
A detailed analysis of the results in Figures~\ref{fig:homo-mosaic-eval} and \ref{fig:mosaic-eval}
can be found in Sections~6.1 and 6.2 of our MICRO 2017 paper~\cite{mosaic}.}

\begin{figure}[h!]
\centering
%  \vspace{-0.5em}%
  \includegraphics[width=\columnwidth]{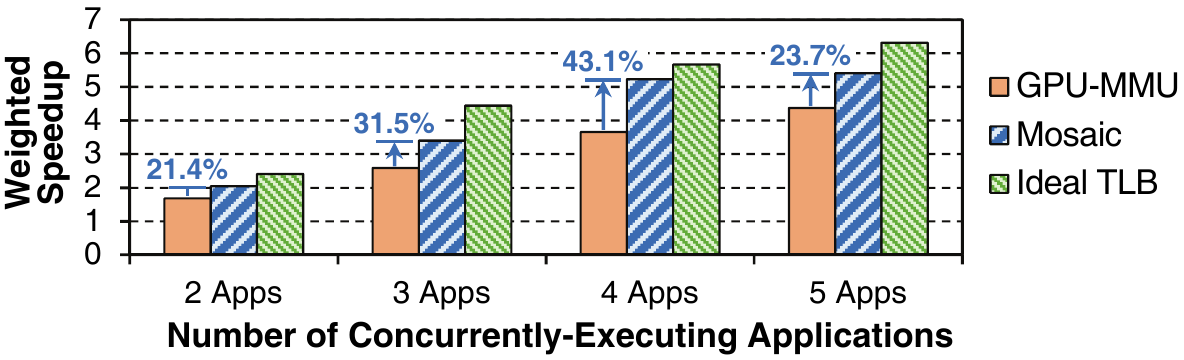}%
%  \vspace{-1em}%
  \caption{Heterogeneous workload performance of the GPU memory managers. Reproduced from~\cite{mosaic}.}%
%  \vspace{-0.5em}%
  \label{fig:mosaic-eval}
\end{figure}

\paragraphbe{Impact of Demand Paging on Performance.} 
All of our results so far show the
performance of the GPU-MMU baseline and \titleShort when demand paging is \emph{enabled}.
Figure~\ref{fig:paging} shows the normalized weighted speedup of the GPU-MMU
baseline and \titleShort, compared to GPU-MMU \emph{without demand paging}, 
where all data required by an application is moved to the GPU
memory \emph{before} the application starts executing.
We make two
observations from the figure. First, we find that \titleShort outperforms GPU-MMU without
demand paging by 58.5\% on average for homogeneous workloads and 47.5\% on average for heterogeneous workloads. 
Second, we find that demand paging has little impact on the weighted speedup.
This is because demand paging latency occurs only when a kernel launches, at which
point the GPU retrieves data from the CPU memory.  The data transfer overhead is
required regardless of whether \changesII{or not} demand paging is enabled, and thus the GPU incurs
similar overhead with and without demand paging. \changesII{We conclude that \titleShort improves
performance significantly, regardless of the demand paging overhead in the baseline.}

\begin{figure}[h]
\centering
%  \vspace{0.5em}%
  \includegraphics[width=\columnwidth]{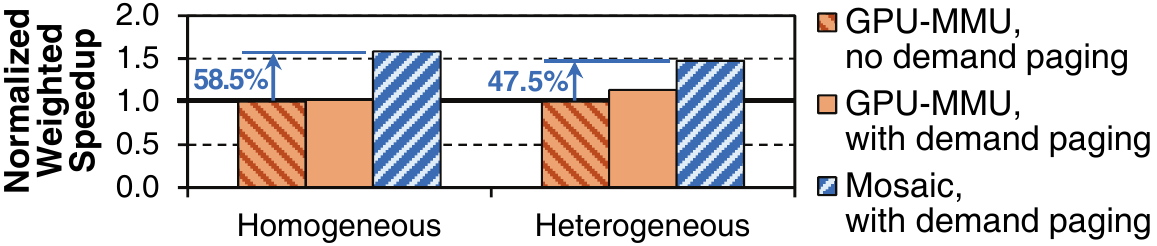}%
%  \vspace{-1em}%
  \caption{Performance of GPU-MMU and \titleShort compared to GPU-MMU without demand paging. Reproduced from~\cite{mosaic}.}%
%  \vspace{-0.5em}%
  %\vspace{-0.5em}%
  \label{fig:paging}
\end{figure}

\textbf{TLB Hit Rate.}
Figure~\ref{fig:mosaic-eval-tlb}
compares the overall TLB hit rate of GPU-MMU to \titleShort{}
for 214 of our 235~workloads, which suffer from \emph{limited TLB reach}
(i.e., workloads that have an L2 TLB hit rate lower than 98\%).
% \changesIIII{, which contains 214 workloads in total \changesIIIII{out of 235 workloads}}).
We make two observations from the figure. First, we observe \changesII{that}
\titleShort is very effective at increasing the TLB reach of these workloads.
%While GPU-MMU is unable to coalesce many base pages into large pages for these
We find that for the GPU-MMU baseline, \emph{every} fully-mapped 
large page frame contains pages from \emph{multiple} applications, as the GPU-MMU allocator
does \emph{not} provide the soft guarantee of \allocatorName (\changesII{i.e., a single large 
page frame contains base pages from only a \emph{single} application}). As a result, GPU-MMU does not have
any opportunities to coalesce base pages into a large page \emph{without} performing
significant amounts of data migration.
In contrast, \titleShort can coalesce a vast majority of base pages thanks to \allocatorName.
As a result, \titleShort reduces the TLB miss rate \changesII{drastically} for these workloads,
with the average miss rate falling below 1\% in both the L1 and L2 TLBs.
% resulting in a massive TLB miss reduction down to 1\% on average in both the L1
% and shared TLBs. 
Second, we observe an increasing amount of interference in
GPU-MMU when more than three applications are running concurrently.
This results in a lower TLB hit rate as the number of applications increases from three to four,
and from four to five.
The L2 TLB hit rate \changesII{of GPU-MMU} drops from 81\% in workloads with two concurrently-executing applications
to 62\% in workloads with five concurrently-executing applications. 
%TLB shootdowns lead to a 3.1\% performance overhead.
\titleShort experiences no such drop due to interference as we increase
the number of concurrently-executing applications, since it makes
much greater use of large page coalescing and enables a much larger
TLB reach. \changesII{We conclude that \titleShort is very effective in improving the hit rate}. 

\begin{figure}[ht!]
\centering
%  \vspace{-0.5em}%
  \includegraphics[width=\columnwidth]{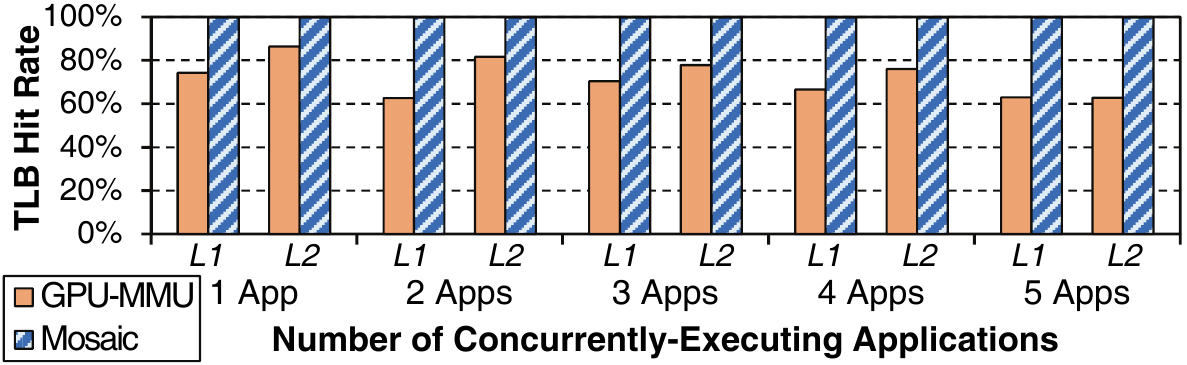}%
%  \vspace{-1em}%
  \caption{L1 and L2 TLB hit rates for GPU-MMU and \titleShort. Reproduced from~\cite{mosaic}.}
%  \vspace{-0.5em}%
  %\vspace{-0.5em}%
  \label{fig:mosaic-eval-tlb}
\end{figure}

We provide the following additional results in \changesII{our full MICRO 2017} paper~\cite{mosaic}:

\begin{itemize}
\item Individual applications' performance with \titleShort and the baseline GPU-MMU
\item TLB size sensitivity of \titleShort and the baseline GPU-MMU
\item Analysis of the effectiveness of \compactionName to reduce memory fragmentation incurs by using large pages
\end{itemize}

\section{Related Work}

To our knowledge, this is the first work to
(1)~analyze the fundamental trade-offs between TLB reach, demand paging 
performance, and internal page fragmentation; and 
(2)~propose an application-transparent GPU memory manager that preemptively 
coalesces pages at allocation time to improve address translation performance, 
while avoiding the demand paging inefficiencies and memory copy overheads 
typically associated with large page support.
% \changesII{To our knowledge, this is the first work that provide a new GPU
% memory management scheme that achieves the benefit of both small and large page sizes.}
Reducing performance degradation from address translation overhead is an
active area of work for CPUs, and the performance loss that we observe as a 
result of address translation is well
corroborated~\cite{direct-segment, jayneel-micro14, bhattacharjee-hpca11,gaud14atc,
merrifield-2016-vee}.  In this section, we discuss previous techniques that aim
to reduce the overhead of address translation \changesII{and demand paging}. % in CPUs, GPUs and heterogeneous
%CPU-GPU architecture.
%\sg{no statement about our novelty here?}
%Gaud et al. observe that large page sizes can harm performance on NUMA machines with
%increased page migration latency and page-level false sharing~\cite{gaud14atc}.

%\subsection{Multiple Page Sizes}
\subsection{TLB Designs for CPU Systems}

TLB miss overhead can be reduced by (1)~accelerating page table
walks~\cite{barr-isca10,large-reach} or reducing the walk
frequency~\cite{jayneel-isca16}; or (2)~reducing the number of TLB misses (e.g.,
through prefetching~\cite{bhattacharjee-pact09, kandiraju-isca02,
saulsbury-isca00}, prediction~\cite{prediction-tlb}, structural changes to
the TLB~\cite{talluri-asplos94, binh-colt, binh-hpca14} or \changesII{a} TLB
hierarchy~\cite{bhattacharjee-hpca11, lustig-13, srikantaiah-micro10,
ahn-tocs15, ahn-isca12, rmm, direct-segment, jayneel-micro14}).

\textbf{Support for Multiple Page Sizes.} Multi-page mapping 
techniques~\cite{talluri-asplos94, binh-colt, binh-hpca14}
use a single TLB entry for multiple page translations,
improving TLB reach by a small
\changesII{factor. Much} greater improvements to TLB reach are needed to
deal with modern memory sizes. 
MIX TLB~\cite{cox17asplos} accommodates entries that translate
multiple page sizes, eliminating the need for a dedicated set of
large page entries \changesII{in the TLB}. MIX TLB is orthogonal to our work, and can be used with \titleShort to
further improve TLB reach.

Navarro et al.~\cite{superpage} identify contiguity-awareness and fragmentation
reduction as primary concerns for large page management, proposing
reservation-based allocation and deferred promotion (i.e., coalescing) of base
pages to large pages.  Similar ideas are widely used in modern OSes~\cite{thp}.
Instead of the reservation-based scheme, Ingens~\cite{ingens} employs a
utilization-based scheme that uses a bit vector to track spatial and temporal
utilization of base pages.

\sloppypar{
\textbf{Techniques to Increase Memory Contiguity.}
%A number of related works propose hardware support to recover and expose memory
%contiguity. 
GLUE~\cite{binh-micro15} groups contiguous, aligned base page
translations under a single speculative large page translation in the TLB.
GTSM~\cite{du15hpca} provides hardware support to leverage the contiguity of
physical memory region even when pages have been retired due to bit errors.
These mechanisms for preserving or recovering contiguity are orthogonal to 
the contiguity-conserving allocation we propose for \titleShort, and they can
help \titleShort by avoiding the need for compaction.
}
\sloppypar{
Gorman et al.~\cite{gorman-ismm08} propose a placement policy for an OS's 
physical page allocator that mitigates fragmentation and promotes contiguity 
by grouping pages according to the amount of migration required
to achieve contiguity. Subsequent work~\cite{gorman-wiosca10}
proposes a software-exposed interface for applications to explicitly
request large pages like {\tt libhugetlbfs}~\cite{libhugetlbfs}.
These ideas are complementary to our work.
\titleShort can potentially benefit from similar policies if \changesII{such policies} can be simplified
enough to be implementable in hardware. 
}

\sloppypar{
\textbf{Alternative TLB Designs.}
Research on shared last-level TLB
designs~\cite{inter-core-tlb, bhattacharjee-hpca11, lustig-13} and page walk cache
designs~\cite{large-reach} has yielded mechanisms that accelerate
multithreaded CPU applications by sharing translations between cores. 
SpecTLB~\cite{spectlb} provides a technique that predicts address
translations. While speculation works on CPU applications,
speculation for highly-parallel GPUs is more complicated~\cite{jablin-pact14,osp-isca13}, and can \changesII{potentially}
waste off-chip DRAM bandwidth, which is a highly-contended resource in GPUs.
Direct segments~\cite{direct-segment}
and redundant memory mappings~\cite{rmm} provide virtual memory support
for server workloads that \changesII{reduces} the overhead of address translation. 
%Both are active areas of research. 
These proposals map large contiguous chunks of virtual memory
to the physical address space in order to reduce the address translation
overhead. While these techniques improve the TLB reach, % in a similar manner as large pages, 
they increase the transfer latency depending on the size of the
virtual chunks they map.
}

\subsection{TLB Designs for GPU Systems}

\textbf{TLB Designs for Heterogeneous Systems.}
Previous works provide several TLB designs for
heterogeneous systems with
GPUs~\cite{powers-hpca14,pichai-asplos14,abhishek-ispass16} and with 
accelerators~\cite{cong-hpca17}. 
\titleShort{} improves upon a state-of-the-art TLB
design~\cite{powers-hpca14} by providing application-transparent,
high-performance support for multiple page sizes in GPUs.
No prior work provides such support.

\textbf{TLB-Aware Warp Scheduler.} Pichai et al.~\cite{pichai-asplos14}
extend the cache-conscious warp scheduler~\cite{ccws} to be aware of the 
TLB in heterogeneous CPU-GPU systems. \changesII{Other more sophisticated warp
schedulers~\cite{ccws,largewarps,warpsub,zheng-cal,caws-pact14,tor-micro13,nmnl-pact13} can also be extended to be TLB aware.}
These techniques are orthogonal to the problem we focus on, and can be applied
in conjunction with \titleShort{} to further improve performance.

\textbf{TLB-Aware Memory Hierarchy.} Ausavarungnirun et al.~\cite{mask}
improve the performance of the GPU under the presence of memory protection by
redesigning the GPU main memory hierarchy to be aware of TLB-related memory
requests. Many prior works propose memory scheduling designs for
GPUs~\cite{complexity,medic,jeong2012qos,adwait-critical-memsched} and
heterogeneous systems~\cite{sms,usui-dash}. These memory scheduling
design can be modified to be aware of TLB-related memory requests and used in
conjunction with \titleShort to further improve the performance of the GPUs.

\textbf{Analysis of Address Translation in GPUs.}
Vesely et al.~\cite{abhishek-ispass16} analyze support for virtual memory in
heterogeneous systems, finding 
that the cost of address translation in GPUs is an order of magnitude
higher than that in CPUs. They observe that high-latency address translations limit the GPU's
latency hiding capability and hurt performance.
Mei et al.~\cite{gpu-arch-microbenchmarking} use a set of microbenchmarks to evaluate the address
translation process in commercial GPUs.
Their work concludes that previous NVIDIA architectures~\cite{kepler,maxwell}
have \emph{off-chip} L1 and L2 TLBs, which lead
to poor performance.

\textbf{GPU Core Modifications.} Many prior
works propose modifications to the GPU core
\changesVIII{design~\cite{caba,vijaykumar.book16,adwait-critical-memsched,gpu-regfile-mohammad,medic,ccws,nmnl-pact13,cpugpu-micro,kuo-throttling14,largewarps,warpsub,zheng-cal,caws-pact14,tor-micro13,ltrf-asplos18}.}
These techniques are complementary to \titleShort, and can be combined with \titleShort to further
improve GPU performance.

\subsection{GPU Virtualization}

VAST~\cite{vast} is a software-managed virtual memory space for
GPUs. In that work, the authors observe that the limited size of physical
memory typically prevents data-parallel programs from utilizing GPUs. To
address this, VAST automatically partitions GPU programs into chunks that fit within the
physical memory space to create \changesII{the illusion of infinite} virtual memory. 
Unlike \titleShort{}, VAST is unable to provide memory protection from 
concurrently-executing GPGPU applications.
Zorua~\cite{zorua} is a holistic mechanism to virtualize
multiple hardware resources within the GPU. Zorua does not virtualize the
main memory, and is thus orthogonal to our work. 
\changesVIII{vmCUDA~\cite{vmCUDA} and rCUDA~\cite{rcuda} provide close-to-ideal
performance, but they require significant modifications to 
GPGPU applications and the operating system, which sacrifice transparency to the application,
performance isolation, and compatibility across multiple GPU architectures.}

\subsection{Demand Paging for GPUs}

Demand paging is a challenge for GPUs~\cite{abhishek-ispass16}.  Recent
works~\cite{tianhao-hpca16, cc-numa-gpu-hpca15}, and the
AMD hUMA~\cite{huma} and NVIDIA PASCAL
architectures~\cite{tianhao-hpca16,pascal}
provide various levels of support for demand paging in GPUs.
These techniques do \emph{not} tackle the existing trade-off in GPUs
between using large pages to improve address translation and
using base pages to minimize demand paging overhead, which
we relax with \titleShort.
%As we discuss in
%Section~\ref{sec:motivation-mosaic}, support for multiple page sizes can be
%adapted to minimize the overhead of demand paging by limiting demand paging to
%base pages only.

\section{Potential Impact of Mosaic}

While several previous works propose mechanisms to lower the overhead of
virtual memory~\cite{direct-segment, jayneel-micro14, bhattacharjee-hpca11,gaud14atc,
merrifield-2016-vee,powers-hpca14,pichai-asplos14,abhishek-ispass16,binh-micro15,du15hpca}, only a handful of these works extensively evaluate
virtual memory on GPUs~\cite{powers-hpca14,pichai-asplos14,abhishek-ispass16,vast}, and no work has investigated
virtual memory as a shared resource when \emph{multiple} GPGPU applications need
to share \changesII{the} GPUs.  In this section, we explore the potential future impact of \titleShort.

\textbf{Support for Concurrent Application Execution in GPUs.} 
The large number of cores within a contemporary GPU make it an attractive
substrate for executing multiple applications in parallel.  This can be especially
useful in virtualized cloud environments, where hardware resources are safely
partitioned across multiple virtual machines to provide efficient resource sharing.
Prior approaches to execute multiple applications concurrently on a GPU have
been limited, as they either (1)~lack sufficient memory protection support across
multiple applications; 
(2)~incur a high performance overhead to provide memory protection; or
(3)~perform a conservative static partitioning of the GPU, 
which can often underutilize many resources in the GPU.

\titleShort provides the first flexible support
for memory protection within a GPU, allowing applications to dynamically partition
GPU resources \changesVIII{\emph{without}} violating memory protection guarantees.
This support can enable the practical virtualization and sharing of GPUs in
a cloud environment, which in turn can increase the appeal of GPGPU programming \changesVIII{and the use cases of GPGPUs}.
By enabling practical support for concurrent application execution \changesVIII{on GPUs},
\titleShort encourages \changesVIII{and enables} future research in several areas, such as
resource sharing mechanisms, kernel scheduling, and quality-of-service
enforcement within the GPU \changesVIII{and heterogeneous systems}.

% \titleShort's support for
% memory protection allows the GPU, \emph{for the first time}, to dynamically
% execute multiple GPGPU applications. Scheduling of multiple GPGPU applications
% can potentially increase the overall GPU utilization and system throughput.
% This enables the opportunity for new techniques that incorporate scheduling at
% multiple granularities (thread block, kernel, application) to further increase
% the performance of GPGPU application. Additional techniques that perform both
% temporal scheduling (i.e.,~\cite{warp-slicer,mafia}) and spatial scheduling
% (i.e.~\cite{mosaic,mask}) of GPGPU applications can provide quality-of-service
% guarantee for system with GPUs without sacrificing performance.

\textbf{Virtual Memory for SIMD Architectures.} \titleShort is an important
first step to enable \changesVIII{\emph{low overhead virtual memory}} in GPUs. % \titleShort enables multiple other research topics. 
%Typically, GPGPU
%applications' performance are generally bounded by memory bandwidth rather than
%long latency memory instructions~\cite{ccws,tbc,tor-micro13,largewarps} as the
%SIMT execution model allows the GPU to hide these long latency memory
%operations effectively. Based on our findings, we uncover that page walks can
%significantly hinder the latency hiding capability and creates a new bottleneck
%in GPU-based systems. We found that a large number of applications' threads
%can share a single page. As a result, a single TLB lookup can stall a large
%number of GPU warps. 
We believe that the key ideas and general observations that we make are applicable to any highly-parallel
SIMD architecture~\cite{flynn}, and to heterogeneous systems with SIMD-based
processing
cores~\cite{bobcat,sandybridge,amd-fusion,apu,kaveri,haswell,amdzen,skylake,powervr,arm-mali,tegra,tegrax1}.
Future works can expand upon our findings and adapt our mechanisms to reduce the overhead of
page walks and demand paging on other SIMD-based systems.

\textbf{Improved Programmability.} Aside from memory protection, virtual
memory can be used to (1)~improve the programmability of GPGPU applications,
and (2)~decouple a GPU kernel's working set size from the size of the GPU
memory. In fact, \titleShort{} transparently allows applications to benefit
from virtual memory without incurring a significant performance overhead. 
This is a key advantage for programmers, many of whom are used to the
conventional programming model used in CPUs to provide application
portability and memory protection.
By providing programmers with a familiar \changesVIII{and simple} memory abstraction, we expect that
a greater number of programmers will start writing \changesVIII{high-performance} GPGPU applications.
Furthermore, by enabling low-overhead memory virtualization, \titleShort{} can
enable new classes of GPGPU applications.  For example, in the past,
programmers were not \changesVIII{able to easily} write GPGPU applications whose memory working set
sizes exceeded the physical memory within the GPU.  With \titleShort,
programmers no longer need to restrict themselves to applications whose working 
sets fit within the physical \changesVIII{memory; they can rely on the GPU itself software-transparently 
managing page migration and address translation.}
% The
% programmability improvement enables various type of applications, which was not
% able to benefit from using the GPU because (1)~they are complicated to be
% programmed for GPUs or (2)~their working set size exceed that of GPU memory, to
% be run on high throughput GPUs.

% \paragraphbe{GPU Resource Utilization.} \titleShort provides a framework to
% enforce memory protection and allow applications to dynamically share the
% GPUs~\cite{mosaic}. % The release of \titleShort
% %, which includes functionality to allow multiple GPGPU applications to execute concurrently, enables
% Future research in the area of GPU resource sharing can further improve the performance
% of GPGPU applications through an intelligent use of resources throughout the
% entire memory hierarhcy. %The design of
% %\titleShort, combined with previously proposed time-share techniques
% %(i.e.,~\cite{warp-slicer,mafia}), allows future research into multiple areas,
% Techniques such as kernel scheduling, resource management, and quality-of-service
% enforcement can be applied to partition GPU cores, caches, TLBs, the main memory,
% and the I/O bus based on each application's characteristics and requirements.
% %All of these cannot future research areas are impossible without providing memory protection. 

\textbf{Publicly-Released Infrastructure.}
Our \changesVIII{simulation} infrastructure is publicly available \changesII{as open-source software~\cite{mosaic.github}}. 
Other researchers can utilize our infrastructure to conduct future research on
virtual memory management on GPUs. Some examples of research topics that can be investigated \changesII{using}
our infrastructure include (1)~how to manage which pages reside in
CPU memory or GPU memory, (2)~how to dynamically partition the physical main memory across multiple
concurrently-executing applications, and (3)~how to maintain programmability of the
virtual memory as the GPU architecture \changesVIII{evolves and becomes more heterogeneous} over time.
We hope and believe that our \changesVIII{new, open-source} infrastructure can inspire future research in these 
and other research areas on GPU \changesVIII{and heterogeneous system} memory virtualization.

\section{Conclusion}

We introduce \titleShort, a new GPU memory manager that provides 
application-transparent support for multiple page sizes.
The key idea of \titleShort is to perform demand paging using \emph{smaller} page
sizes, and then coalesce small (i.e., base) pages into a \emph{larger} page
immediately after allocation, which allows address translation to use large 
pages and thus increase TLB reach.
We have shown that \titleShort significantly outperforms state-of-the-art GPU 
address translation designs and achieves performance close to an ideal TLB, 
across a wide variety of workloads. We conclude that \titleShort effectively combines 
the benefits of large pages and demand paging in GPUs, thereby breaking the 
conventional tension that exists between these two concepts. 
We hope the ideas presented in our MICRO 2017 paper can lead to future works that analyze 
\titleShort in detail and provide even lower-overhead support for synergistic 
address translation and demand paging in \changesVIII{GPUs and} heterogeneous systems.

\section*{Acknowledgments}
We thank the anonymous reviewers and SAFARI group members for
their feedback. Special thanks to Nastaran Hajinazar, Juan Gómez
Luna, and Mohammad Sadr for their feedback. We acknowledge
the support of our industrial partners, especially Google, Intel,
Microsoft, NVIDIA, Samsung, and VMware. This research was
partially supported by the NSF (grants 1409723 and 1618563), the
Intel Science and Technology Center for Cloud Computing, and the
Semiconductor Research Corporation.

{
\bibliographystyle{IEEEtranS}
\bibliography{references}
}

\end{document}